# Defining, Estimating and Using Credit Term Structures

## Part 3: Consistent CDS-Bond Basis

**Arthur M. Berd**
Lehman Brothers Inc.

**Roy Mashal**
Lehman Brothers Inc.

**Peili Wang**
Lehman Brothers Inc.

*In the third part of this series we introduce consistent relative value measures for CDS-Bond basis trades using the bond-implied CDS term structure derived from fitted survival rate curves. We explain why this measure is better than the traditionally used Z-spread or Libor OAS and offer simplified hedging and trading strategies which take advantage of the relative value across the entire range of maturities of cash and synthetic credit markets.*

### INTRODUCTION

In a recent paper (Berd, Mashal and Wang [2004a], cited hereafter as Part 1) we introduced a new methodology for direct estimation of implied term structures of survival probabilities from credit bond prices. We have shown that this methodology is more robust than the traditional implementations of reduced-form default models (for the latter, see Jarrow and Turnbull [1995], and Duffie and Singleton [1999]). More importantly, it is more consistent with the underlying bankruptcy resolution practices such as debt acceleration and equal priority recovery for the same-seniority bonds.

Our methodology is well suited to a direct comparison with credit derivatives, particularly credit default swaps, whose valuation is driven by the modeling of default probabilities. In this paper we introduce new relative value measures, which take advantage of the internal consistency of this pricing methodology. In particular, we define:

- Bond-Implied CDS (BCDS) term structure
- CDS-Bond curve basis
- Systematic and full bond-specific basis to CDS curve
- Risk-free-equivalent coupon (RFC) streams for credit-risky bonds

We also introduce and discuss static replication/hedging strategies of credit risk in cash bonds using forward and spot CDS. In particular, we demonstrate in detail how these strategies can be used to hedge the default risk of a credit bond with an arbitrary coupon, given an arbitrary term structure of risk-free rates. The complete hedge of credit risk in these strategies is reflected in the complementarity between the risk-free-equivalent coupon streams and the CDS hedging costs, which is formally proven in the Appendix.

### BOND-IMPLIED CDS TERM STRUCTURE

Credit default swaps are by far the largest component of the rapidly growing credit derivatives market. They comprise as much as 70% of the market by notional volume. The outstanding notional of CDS is comparable with that of cash bonds, and their liquidity often exceeds that of the cash market for the top 200 or so names. At the same time, cash credit bonds often cover a greater range of maturities than CDS and have far more extensive historical data associated with them, providing fertile ground for research and back-testing trading strategies. One can therefore hope that an implied CDS measure derived from bond prices will be a valuable tool for consistent comparisons between the two markets across the entire range of maturities.

### Deriving the Bond-Implied CDS spread term structure

The survival-based valuation approach is well suited to the CDS market. In fact it has been the market practice since its inception. By deriving the bond-implied CDS spreads within the same framework we are aiming to give investors an apples-to-apples relative value measure across the bond and CDS markets.

The pricing relationship for credit default swaps simply states that the expected present value of the premium leg is equal to the expected present value of the contingent payment (see Duffie [1999], Hull and While [2001] and O'Kane and Turnbull [2003]):

$$[1] \quad \frac{S}{f} \cdot \sum_{i=1}^{N} Q(t,t_i) \cdot Z(t,t_i) = (1-R) \cdot \sum_{i=1}^{N} \left( Q(t,t_{i-1}) - Q(t,t_i) \right) \cdot Z(t,t_i)$$

where, $S$ is the annual spread, $f$ is the payment frequency ($f=4$), $Q(t,t_i)$ is the issuer's survival probability between time $t$ to time $t_i$, $Z(t,t_i)$ stands for the (default risk-free) discount factor from time $t$ to time $t_i$, and $R$ stands for the recovery rate. This formulation is consistent with the methodology described in Part 1 of this series.

The bond-implied CDS spread term structure, hereafter denoted as BCDS term structure, is defined by substituting the survival probability term structure fitted from bond prices, $Q_{bond}(t,t_i)$, into the following equation for par CDS spreads:

$$[2] \quad BCDS(t,t_N) = f \cdot (1-R) \cdot \frac{\sum_{i=1}^{N} \left( Q_{bond}(t,t_{i-1}) - Q_{bond}(t,t_i) \right) \cdot Z(t,t_i)}{\sum_{i=1}^{N} Q_{bond}(t,t_i) \cdot Z(t,t_i)}$$

The different payment frequencies for bonds and CDS do not represent a problem because the fitted survival probability term structure is continuous and can be evaluated at any frequency. Furthermore, the CDS contracts stipulate that at default any accrued protection premium must be paid. The latter feature of the CDS contract makes the following continuous-time approximation quite accurate and useful:

$$[3] \quad BCDS(t,T) = (1-R) \cdot \frac{\int_t^T du \cdot h(u) \cdot e^{-\int_t^u (r(s)+h(s))ds}}{\int_t^T du \cdot e^{-\int_t^u (r(s)+h(s))ds}}$$

For a more accurate evaluation of the CDS spread given the term structure of the hazard rates see O'Kane and Turnbull (2003).

### Comparison with conventional spread measures

The BCDS term structure complements bond-based valuation measures defined in Part 1, namely the par Libor and par Treasury spreads (P-spreads), and constant coupon price (CCP) term structures. Under certain circumstances, the BCDS term structure may be significantly different from conventional measures such as a bond's Z-spread and asset swap spread (see O'Kane and Sen [2004] for definitions of the conventional spread measures).

Market participants often use the Z-spread as a proxy for comparing bonds with CDS. Such analysis may be misleading because the derivation of Z-spreads is based on the valuation of credit bonds with spread-based discount functions which we have shown to be incorrect in our earlier paper. This is due to the fact that credit bonds do not have fixed cashflows – they only have fixed promised cashflows, while realized cashflows may well turn out to be different from the pro-forma projections. Hence, a survival-based approach is needed to correctly model default-risky bonds.

The Z-spread of a credit bond is consistent with a correct survival-based valuation framework only under the assumption of zero recovery rates. Generally speaking, such an assumption is far from observed statistics. Given that historical average recovery values are about 40% (30% during the recent credit downturn) the Z-spread overestimates the losses in case of default by a significant amount. Therefore, it should not be surprising that the BCDS spread term structure can differ from the Z-spread by substantial margins.

An example of this is presented in Figures 1a and 1b, which contrasts the survival-based BCDS curve with the Libor spread curve and the Z-spreads of individual bonds, based on spread-based discount function methodology. Figure 1a shows the results for Georgia Pacific Co. as of December 31, 2002 – ie, at a time when the bonds of the company were substantially distressed. We observe that the shape of the BCDS curve bears no resemblance to the shape of the traditionally fitted Libor spread curve or to the Z-spreads of individual bonds. Figure 1b, on the other hand, shows the same set of bonds as of December 31, 2003 – when the spreads have generally tightened and the bonds no longer trade at large discounts. Here, the BCDS curve is much closer to the conventional Libor spreads.

We would like to emphasize that the methodology developed in Berd *et al.* (2003) is based on fitting the *prices* of bonds rather than their spreads, and that the deviation of the BCDS curve from bond Z-spreads does not indicate a poor fit of bond values. To confirm this, we also show the constant coupon price term structures (ie, the projected prices of hypothetical bonds with 6%, 8% and 10% coupons) from the same model compared with the bond prices as of December 31, 2002. As we can see, the CCP term structures neatly envelop the scatter plot of bond prices, as indeed they should.

Figures 1a and 1b also shed light on the interesting issue regarding the "slope of the spread curve". In both cases, the Libor OAS curve which is fitted using the conventional methodology, is inverted, while the BCDS curve is not (or at least not for all maturities). One often hears that distressed bond pricing is always accompanied by an inverted spread curve. However, this is only true in the conventional, spread-discount-factor based models. As discussed in great detail in the Part 1 of this series, the spread inversion in these models is largely a consequence of their inability to correctly capture the peculiar aspects of distressed bond pricing – namely the fact that such bonds trade "on price" or "to recovery". Unlike the Z-spreads, the BCDS term structure need not be necessarily inverted in order to capture the high credit risk of distressed bonds.

**Figure 1a.    BCDS and Libor OAS term structures and bond Z-spreads, GP as of 12/31/02**

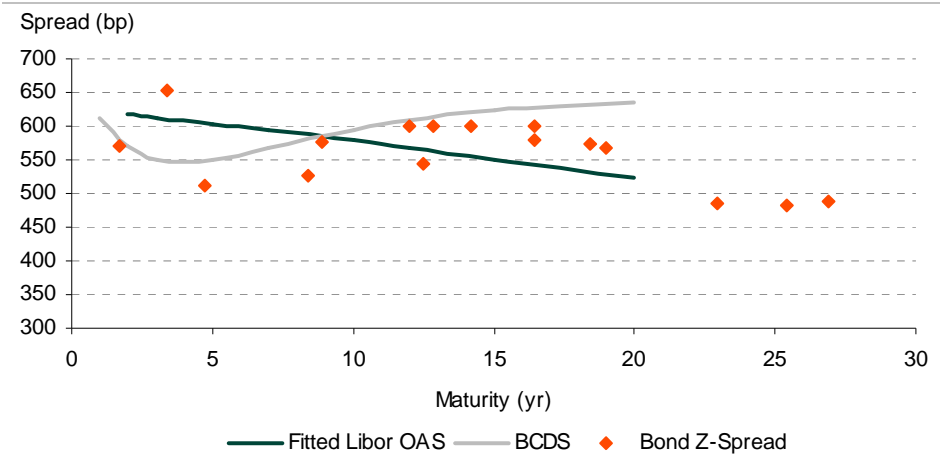

**Figure 1b.    BCDS and Libor OAS term structures and bond Z-spreads, GP as of 12/31/03**

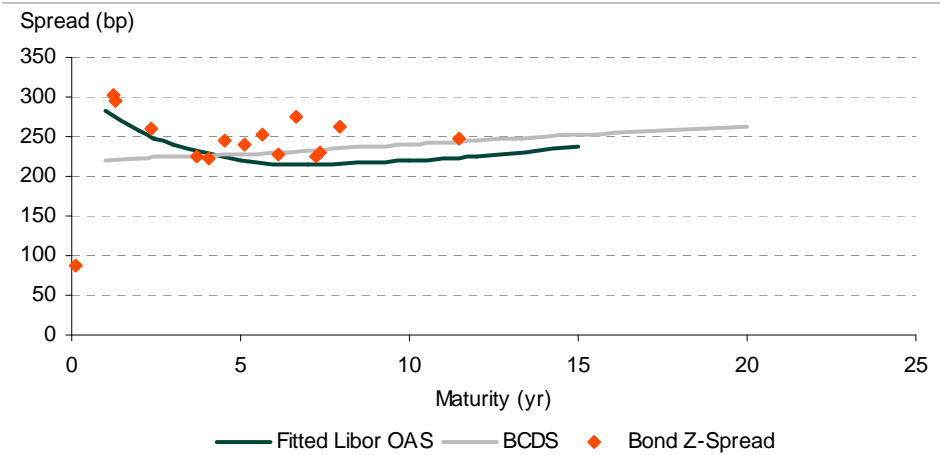

**Figure 1c.    CCP term structures and bond prices, GP as of 12/31/02**

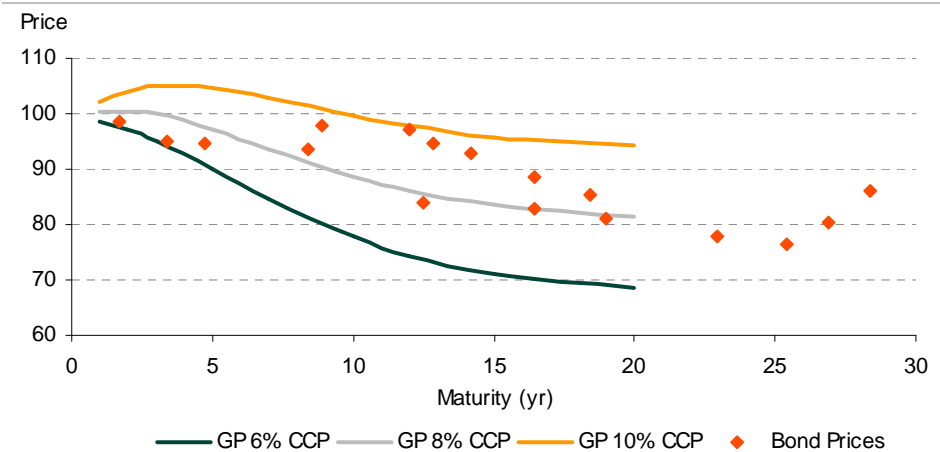

## STATIC HEDGING OF BONDS WITH CDS

In the previous section we defined the BCDS term structure and explained why it is a better valuation measure than the traditional Libor OAS or Z-spread measures. However, in order to be sure that the BCDS term structure corresponds to the fair value of the CDS as seen by the bond market, one should explicitly construct a hedging strategy using CDS that would completely eliminate the credit risk of any given cash bond.

Such a strategy is well known for a par bond when the underlying credit risk-free curve (which is usually assumed to coincide with the Libor swaps curve) and the hazard rates are flat. Under such assumptions, the hedging strategy turns out to be simple – buy the credit bond and buy an equal face value of CDS. The resulting combination exhibits no risk due to default. The interest rate risk of the hedged position coincides with the risk of a credit risk-free bond with a lower coupon equal to the difference of the credit-risky coupon and the CDS spread. This risk can be hedged using interest rate swaps, although residual timing risk may remain if the swap is not terminated upon default.

The situation quickly becomes more complex when the above assumptions are not valid, most notably in case of non-par bonds and/or a Libor curve that is significantly different from flat. Some of the difficulties and possible remedies have been discussed in McAdie and O'Kane (2001). There it was shown that, when hedging a given fixed coupon bond with a CDS of matching maturity, neither the face value hedge ratio nor the market value hedge ratio based on the current price of the bond is satisfactory, because they lead either to a residual mark-to-market risk during the life of the bond or to a substantial dependence of the carry cost of the hedged position on the price of the bond. McAdie and O'Kane conclude that, for a hedge using a single CDS position, an intermediate strategy called a zero-recovery market value hedge provides the best compromise.

In this section we further extend the intuition put forward in McAdie and O'Kane (2001). As it turns out, the complete hedging strategy cannot in general be accomplished by a single CDS position. Indeed, under generic conditions, a bond will have a non-trivial term structure of forward prices computed with today's risk-free discount curve and hazard rate term structure. If a bond is well hedged today, it will likely be either over- or under-hedged in the future, when its price is expected to change. After all, if the bond actually survives to maturity its price will pull to par. So the hedge at maturity should always be based on the expected price at that point, which is close to par. The question is – what should happen at prior times?

### Staggered hedging strategy with forward CDS

The precise answer to this question is presented in the Appendix, where we define the notion of the *risk-free-equivalent coupon stream* (RFC) and show its complementarity with the BCDS curve. The complementarity condition defines the unique static hedging strategy based on a sequence of forward CDS contracts with notional that depends on the bond's forward price for each maturity.

A forward CDS contract provides protection during a future time period in exchange for premiums which are paid during that period but whose level is preset today. In case of a credit event prior to the starting date of that period, the forward CDS knocks out without a payment and provides no further protection. However, if a credit event occurs during the contractual future period – the CDS pays the loss amount $N_{fwd} \cdot (1-R)$. Since, based on today's calculations, the bond is expected to have a forward price $P_{fwd} \neq 100\%$ at that future date, then in order to insure the full forward price of the bond we must choose the hedge notional

$$[4] \quad N_{fwd}(t) = \frac{P_{fwd}(t) - R}{1 - R}$$

This intuition is precisely right, as shown in the Appendix. The sequence of forward CDS hedges does, in fact, completely hedge out the credit risk of a fixed coupon bond. If we subtract from the bond's coupon the cost of hedging according to [4], the residual cashflows will coincide with the above mentioned risk-free-equivalent coupon stream. The RFC is such a pre-set sequence of term-dependent coupons which, if discounted with risk-free rate, correspond to the same term structure of forward prices as the one obtained for the credit-risky bond under consideration (see Appendix for more details).

To be more accurate, we must mention that the hedge notional shown above is only valid for the case of continuously compounded coupons and spreads. In a more realistic case of semi-annual coupon payments, one must take into account the present value of those coupons as well as the fact that the CDS market conventions stipulate a payment of the accrued premium upon default. The result is a slightly modified expression for the hedge notional applicable for the payment period ending at $t_i$:

$$[5] \quad N_{fwd}(t_i) = \frac{0.5 \cdot \left(P_{fwd}(t_{i-1}) + P_{fwd}(t_i) + w \cdot C\right) - R}{1 - R}$$

The hedge depends on the average forward price during the preceding coupon period as well as the potential loss of coupon in case of default. The effective weight $w$ depends on the assumed coupon recovery. We have found empirically that for coupon recovery of 50% the weight is also 50%, while for coupon recovery of 0% the weight is 25%.

Figure 2a illustrates this staggered forward CDS strategy in the case of a hypothetical 8% coupon 5-year bond which trades at a high initial price premium of 116.69%. The top rows of the table show the term to maturity, the spot BCDS curve and the forward BCDS curve with forward horizon equal to 0.5 years (see Berd and Ranguelova [2004] for the relationship between spot and forward CDS spreads). The middle of the table shows how the hedge notionals are related to the forward prices of the credit bond and how they gradually decrease as the forward prices exhibit pull to par – compare the notional for a given term (column) with the forward price shown in the shaded area on the row corresponding to the same term (in this example we used 50% principal and coupon recovery in equation [5]). The next four rows demonstrate the complementarity between the hedging costs and the risk-free-equivalent coupon stream (RFC), which holds very accurately. Finally, we show that the forward prices of the resulting residual cashflow streams discounted with risk-free interest rates do indeed coincide with high accuracy with the forward prices of the original bond for all future horizons.

### Staggered hedging strategy with spot CDS

One could, in principle, construct a very similar strategy using the spot instead of forward CDS. As discussed in Berd and Ranguelova (2004), a forward CDS protection contract is closely related to a long-short pair trade in spot CDS, with equal notional long protection position at the longer maturity and short protection position at the shorter maturity. Although in terms of credit protection the long-short trade is indeed equivalent to a forward CDS contract, the premium legs of these two strategies will generally differ, resulting in different forward mark-to-markets.

If we execute the hedging strategy with long-short pairs, the result becomes a staggered hedge which is nearly 100% notional for the final maturity, and which includes some additional relatively small long (or short) positions for shorter maturities depending on the forward prices of the credit bond being hedged. Each such position hedges the incremental digital price risk (with no recovery) corresponding to the next maturity interval on the hedging grid:

$$[6] \quad N_{pair}(t_i) = \frac{P_{fwd}(t_i) - P_{fwd}(t_{i+1})}{1 - R}$$

Figure 2b shows an implementation of this strategy for the same hypothetical high premium bond. While the current credit risk of this bond is indeed hedged, as evidenced by close replication of spot price, this hedge does drift away from perfection with time (but remains quite close nevertheless). This is because the pair trades are not exactly equivalent to a series of forward CDS.

### The coupon and price premium/discount dependence

From the discussion and examples shown above, it is clear that both the coupon level of the credit bond and the term structure of the underlying interest rates and issuer's hazard rates may substantially affect the hedging strategy with forward CDS or long-short CDS pairs. Its dependence on the underlying bond is depicted in Figures 3a, 3b and 3c.

Figure 3a shows a case of a bond with a high coupon equal to 8% and a high current price of 116.69% – the same as Figure 1a. The forward CDS hedge notional starts as high as 133% of the face value, and gradually decreases toward 100%. Despite the decrease in the hedge notional, the semi-annual cost of hedging grows gradually from 40bp to 44bp as a result of a relatively steep forward CDS curve term structure.

Figure 3b shows a case of a bond with a very low coupon equal to 3% and a current discount price equal to 94.33%. The forward CDS hedge notional starts at 89% of the face value, and gradually increases toward 100%. The semi-annual cost of hedging grows more steeply from 27bp to 43bp as both forward CDS rates and hedge notionals grow.

Figure 3c shows a case of a bond with near-par coupon equal to 4.25% and a current price of 99.93%. Despite the fact that this is a par bond, the forward bond prices and hedge notionals exhibit a non-trivial term structure, starting near 100%, then dropping to lower levels and only pulling back to par near final maturity. The semi-annual cost of hedging grows from 33bp to 43bp, which is somewhere between the high and low coupon cases.

### The residual interest rate risk

The static hedging strategy using CDS addresses only the credit risk exposure. According to the complementarity principle proven in Appendix A, the remainder of these hedging strategies is the risk-free-coupon stream (RFC) bond whose forward price profile matches precisely that of the credit bond. While this hypothetical RFC bond is no longer subject to credit loss risk it is still subject to interest rate risk. In order to fully hedge the residual interest rate risk one would simply have to swap all the projected RFC cash flows into floating rate using a sequence of interest rate swaps of appropriate maturities.

This, however, does not fully eliminate the interest rate risk. Indeed, upon a default event the cash flows from the credit bond itself and all the CDS will terminate, while the interest rate swaps that have maturities longer than the date of default will still be outstanding. The expected net market value of these remaining swaps is equal to the expected variation of the interest rate hedge package from the forward price of the RFC bond. Hence, it is equal to zero by construction regardless of the default timing. In order to fully eliminate not only the expected risk exposure but also the residual risk in all states of the world both before and after default event one would need to use the so-called "perfect asset swaps", i.e. fixed-for-floating swaps which contractually terminate upon the default of the reference credit entity.

**Figure 2a.  Complete hedge of a premium credit bond with forward CDS, 8% coupon 5-year maturity bond**

| | Term | 0.00 | 0.50 | 1.00 | 1.50 | 2.00 | 2.50 | 3.00 | 3.50 | 4.00 | 4.50 | 5.00 |
|---|---|---|---|---|---|---|---|---|---|---|---|---|
| | Fwd BCDS | 0.60% | 0.60% | 0.63% | 0.66% | 0.69% | 0.72% | 0.74% | 0.77% | 0.80% | 0.83% | 0.86% |
| | Spot BCDS | 0.60% | 0.60% | 0.62% | 0.63% | 0.65% | 0.66% | 0.67% | 0.69% | 0.70% | 0.71% | 0.72% |
| **Term** | **Fwd Price** | | | | | | | | | | | |
| 0.00 | 116.70% | | | | | | | | | | | |
| 0.50 | 113.84% | | 1.33 | | | | | | | | | |
| 1.00 | 111.31% | | | 1.27 | | | | | | | | |
| 1.50 | 109.09% | | | | 1.22 | | | | | | | |
| 2.00 | 107.16% | | | | | 1.18 | | | | | | |
| 2.50 | 105.50% | | | | | | 1.15 | | | | | |
| 3.00 | 104.08% | | | | | | | 1.12 | | | | |
| 3.50 | 102.84% | | | | | | | | 1.09 | | | |
| 4.00 | 101.77% | | | | | | | | | 1.07 | | |
| 4.50 | 100.84% | | | | | | | | | | 1.05 | |
| 5.00 | 100.00% | | | | | | | | | | | 1.03 |
| | **Hedge Notional** | | **1.33** | **1.27** | **1.22** | **1.18** | **1.15** | **1.12** | **1.09** | **1.07** | **1.05** | **1.03** |
| | **Protection CF** | | 0.40% | 0.40% | 0.40% | 0.41% | 0.41% | 0.42% | 0.42% | 0.43% | 0.43% | 0.44% |
| | **Coupon less Protection CF** | | 3.60% | 3.60% | 3.60% | 3.59% | 3.59% | 3.58% | 3.58% | 3.57% | 3.57% | 3.56% |
| | **RCF** | | 3.60% | 3.60% | 3.60% | 3.60% | 3.59% | 3.59% | 3.58% | 3.57% | 3.56% | 3.56% |
| | *CF Diff* | | *-0.01%* | *0.00%* | *0.00%* | *0.00%* | *0.00%* | *0.00%* | *0.00%* | *0.00%* | *0.00%* | *0.00%* |
| | **Projected Fwd Price** | 116.69% | 113.83% | 111.30% | 109.09% | 107.16% | 105.51% | 104.08% | 102.85% | 101.77% | 100.84% | 100.00% |
| | **Fwd Price** | 116.70% | 113.84% | 111.31% | 109.09% | 107.16% | 105.50% | 104.08% | 102.84% | 101.77% | 100.84% | 100.00% |
| | *Price Diff* | *-0.02%* | *-0.01%* | *-0.01%* | *0.00%* | *0.00%* | *0.00%* | *0.00%* | *0.00%* | *0.00%* | *0.00%* | *0.00%* |

Figure 2b.  Present value hedge of a premium credit bond with spot CDS, 8% coupon 5-year maturity bond

| Term | | 0.00 | 0.50 | 1.00 | 1.50 | 2.00 | 2.50 | 3.00 | 3.50 | 4.00 | 4.50 | 5.00 |
|---|---|---|---|---|---|---|---|---|---|---|---|---|
| | **Fwd BCDS** | 0.60% | 0.60% | 0.63% | 0.66% | 0.69% | 0.72% | 0.74% | 0.77% | 0.80% | 0.83% | 0.86% |
| | **Spot BCDS** | 0.60% | 0.60% | 0.62% | 0.63% | 0.65% | 0.66% | 0.67% | 0.69% | 0.70% | 0.71% | 0.72% |
| **Term** | **Fwd Price** | | | | | | | | | | | |
| 0.00 | 116.70% | | | | | | | | | | | |
| 0.50 | 113.84% | | 1.33 | | | | | | | | | |
| 1.00 | 111.31% | | -1.27 | 1.27 | | | | | | | | |
| 1.50 | 109.09% | | | -1.22 | 1.22 | | | | | | | |
| 2.00 | 107.16% | | | | -1.18 | 1.18 | | | | | | |
| 2.50 | 105.50% | | | | | -1.15 | 1.15 | | | | | |
| 3.00 | 104.08% | | | | | | -1.12 | 1.12 | | | | |
| 3.50 | 102.84% | | | | | | | -1.09 | 1.09 | | | |
| 4.00 | 101.77% | | | | | | | | -1.07 | 1.07 | | |
| 4.50 | 100.84% | | | | | | | | | -1.05 | 1.05 | |
| 5.00 | 100.00% | | | | | | | | | | -1.03 | 1.03 |
| | | | | | | | | | | | | |
| | **Hedge Notional** | | 0.05 | 0.05 | 0.04 | 0.04 | 0.03 | 0.03 | 0.02 | 0.02 | 0.02 | 1.03 |
| | **Protection CF** | | 0.46% | 0.45% | 0.43% | 0.42% | 0.41% | 0.40% | 0.39% | 0.38% | 0.37% | 0.37% |
| | **Coupon less Protection CF** | | 3.54% | 3.55% | 3.57% | 3.58% | 3.59% | 3.60% | 3.61% | 3.62% | 3.63% | 3.63% |
| | **RCF** | | 3.60% | 3.60% | 3.60% | 3.60% | 3.59% | 3.59% | 3.58% | 3.57% | 3.56% | 3.56% |
| | *CF Diff* | | -0.07% | -0.05% | -0.03% | -0.01% | 0.00% | 0.02% | 0.03% | 0.05% | 0.06% | 0.07% |
| | **Projected Fwd Price** | 116.76% | 113.96% | 111.48% | 109.30% | 107.39% | 105.73% | 104.29% | 103.02% | 101.91% | 100.91% | 100.00% |
| | **Fwd Price** | 116.70% | 113.84% | 111.31% | 109.09% | 107.16% | 105.50% | 104.08% | 102.84% | 101.77% | 100.84% | 100.00% |
| | *Price Diff* | 0.06% | 0.13% | 0.17% | 0.21% | 0.22% | 0.22% | 0.21% | 0.18% | 0.14% | 0.08% | 0.00% |

**Figure 3a.**     **Hedging strategy for high coupon, premium bond**

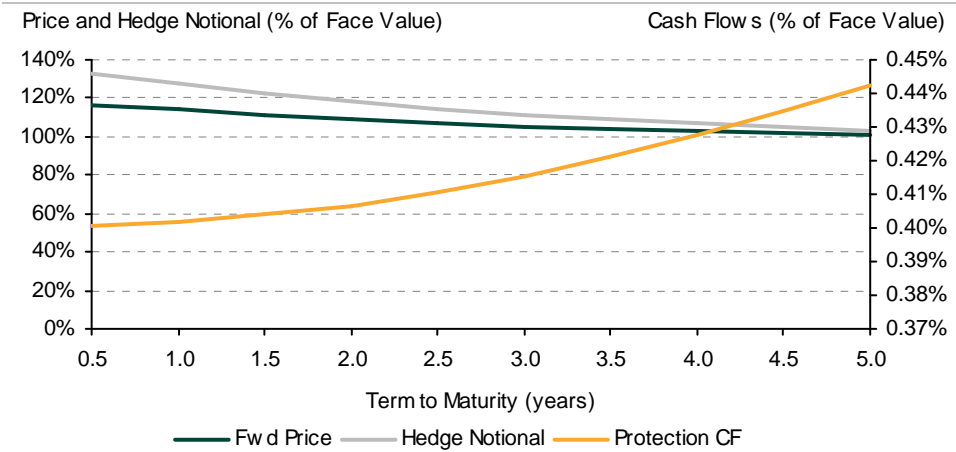

**Figure 3b.**     **Hedging strategy for low coupon, discount bond**

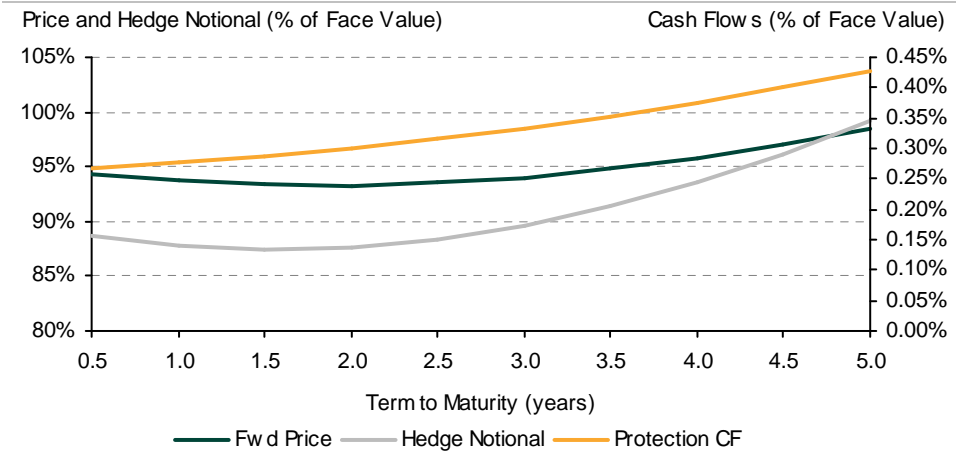

**Figure 3c.**     **Hedging strategy for medium coupon, near-par bond**

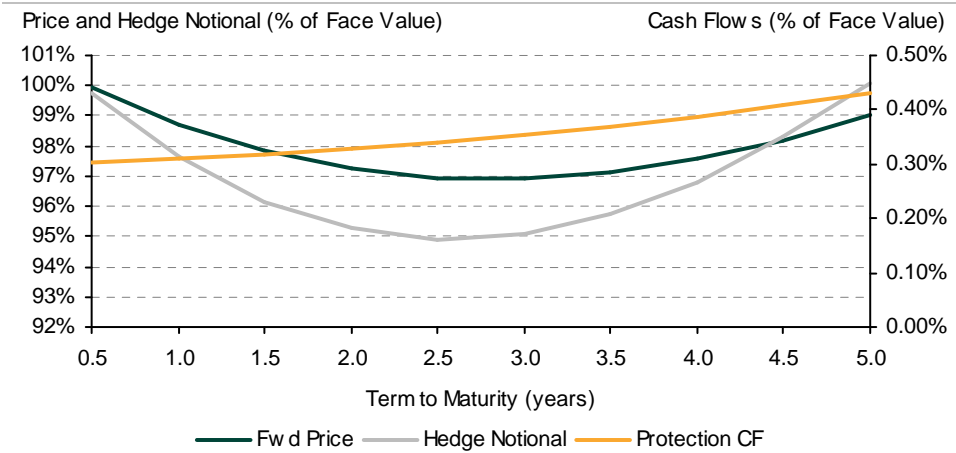

### RELATIVE VALUE MEASURES FOR BASIS TRADING

Although CDS and cash bonds reflect the same underlying issuer credit risk, there are important fundamental and technical reasons why the CDS and bond markets can sometimes diverge from the economic parity (see McAdie and O'Kane [2001]). Such divergences, commonly referred to as the CDS-Bond basis, are closely monitored by many credit investors. Trading the CDS-Bond basis is one of the widely used strategies for generation of excess returns using CDS. To facilitate basis trading, one must have a reliable measure of relative value between cash bonds and CDS. Here we suggest such a measure based on our approach.

#### The CDS-Bond curve basis

If the bonds of a given issuer were perfectly priced according to our framework, their prices would satisfy the survival-based fair value, and the BCDS term structure derived from the issuer hazard rates would correspond to a complete hedge of the credit risk as discussed in the previous section. Therefore, if the market-observed CDS term structure coincided with the BCDS term structure, then one could claim that there is no basis between the two markets since the hedged cash bond would have zero expected excess return over risk-free rates. If the market-observed CDS spreads were tighter than BCDS, one could hedge the credit bond at a cheaper price than that implied by the model, thus locking in a positive expected excess return. Vice versa, if the market CDS spreads were wider than BCDS, the hedge would have a higher cost, and the expected excess return of the hedged position would be negative.

We therefore introduce the **CDS-Bond curve basis** as a consistent relative value measure:

[7] $$\text{CurveBasis}(0,T) = CDS_{market}(0,T) - BCDS(0,T)$$

This measure corrects for the biases associated with commonly used asset swap spreads and Z-spreads. Since this basis measure is well defined across the entire range of maturities, one can use it not only to screen for issuers that correspond to attractive CDS-Bond trading opportunities, but also to pinpoint maturities for which such trades would be most beneficial.

#### Bond-specific basis and incremental relative value

The curve basis does not fully reflect the differences in the costs of hedging implied by the fair value BCDS term structure and by the market observed CDS term structure. These differences depend, as explained in the previous section, on the projected forward price term structure of cash bonds, and therefore on the coupon level of the bond under consideration.

According to the staggered hedging strategy presented in the previous section, the notional of the final maturity hedge using the spot CDS is approximately equal to the face value and therefore its contribution to the CDS-Bond basis is simply the curve basis. However, the incremental hedges for each maturity horizon are proportional to the change in the projected forward price of the bond during the short maturity span around that horizon [6]. The present value of every basis point of difference between the BCDS and the market observed CDS curve for each of these incremental hedges is given by the risky PV01 calculated in accordance with the BCDS curve. Therefore, the incremental hedging cost differential $HCD_{curve}(T)$ due to intermediate maturity curve basis is equal to:

[8]
$$HCD_{curve}(T) = -\int_0^T \text{CurveBasis}(0,t) \cdot \frac{1}{1-R} \cdot \frac{\partial P_{fwd}(t,T)}{\partial t} \cdot \text{RiskyPV01}(0,t) \cdot dt$$

To convert the hedging cost differential into a spread-equivalent measure, we divide it by the risky PV01 of the final maturity BCDS. This has a meaning of a weighted average curve basis in units corresponding to matching maturity CDS. We call it a **systematic bond basis**:

$$[9] \quad \text{BondBasis}_{\text{systematic}}(T) = \text{CurveBasis}(0,T) + \frac{HCD_{curve}(T)}{\text{RiskyPV01}(0,T)}$$

Generally speaking, if the bond price premium or discount is not large, then the corrections to the systematic bond basis due to intermediate maturities will be very small, and one can use the curve basis for the final maturity as a good proxy of the systematic bond basis. However, one must note that if the matching maturity curve basis is exactly zero, then the correction term will become the main component of the systematic bond basis. Its dependence on the bond price and curve basis is non-trivial.

If the curve basis is positive at intermediate maturities, and the bond is trading at premium and is expected to gradually accrete to par, the systematic bond basis will be positive – ie, it will cost more to hedge the bond than the BCDS curve would suggest. If, on the contrary, the curve basis is positive but the bond is trading at a discount and is expected to accrete up to par, then the systematic bond basis will be negative – ie, it will cost less to hedge the bond with CDS than the bond market itself implies. This is because the staggered hedge strategy in this case actually requires selling protection at intermediate maturities, and therefore the positive curve basis will net extra benefits for the hedger. In the other cases when either the curve basis changes sign at intermediate maturities or the bond trades near par and its forward price may both increase and decrease over the future time horizon, the hedging cost differential can easily turn out to be either positive or negative – one would have to perform the full calculation to find out.

In addition to the fair value hedging cost differential embodied in the curve basis, there will also be an issue-specific pricing differential corresponding to the OAS-to-Fit (OASF) measure introduced in Berd *et. al.* (2003). This measure captures the pricing differential between the given bond and the issuer's fitted survival curve. As such, it reflects liquidity and other technical aspects of bond pricing unrelated to credit risk.

We define the **full bond-specific CDS-Bond** basis as the systematic bond basis minus an incremental amount equal to the bond's OASF:

$$[10] \quad \text{BondBasis}_{\text{full}}(T) = \text{BondBasis}_{\text{systematic}}(T) - OASF$$

Depending on the sign and magnitude of OAS-to-Fit, a given bond may have a negative basis while the issuer curve basis is positive, and vice versa. Having said this, since the average OASF across all bonds is zero by construction (see the discussion in our previous paper) then our definition of the bond-specific basis does not add any new bias to the systematic (or curve) basis. It only serves to assist investors in picking the best candidates for execution of either positive or negative basis strategies. For example, if the investor believes that a negative curve basis will converge, then picking the cheapest bond (most positive OASF) will add an incremental expected return to the trade.

### Examples of curve and bond-specific basis

Let us consider an example of curve basis in one of the most actively traded issuers, Altria Group (ticker MO). Using the data as of February 6, 2004 we can see the difference between the shapes of the BCDS and market CDS curves. A clear hump in the relative curve basis in intermediate maturities is related to significant hedging activity in the CDS market as volatility picked up in late January and early February.

If an investor's view is that the curve basis is a transient phenomenon and is expected to converge, then picking a two- to three-year maturity range would maximize the convergence potential. In this maturity range the CDS market looked cheap to cash using our relative value measures (we use the bid-side CDS marks since the bond valuation measures including BCDS term structure are also based on bid-side quotes).

On a bond-by-bond comparison the 5-year MO 5.625 of 2008 (P=104.12) has the largest negative OAS-to-Fit and is therefore the richest cash bond among those included in this analysis. However, when considering the combined BCDS+OASF measure, the largest total basis is exhibited by the 2-year bonds MO 6.375 of 2006 (P=106.07) and MO 6.95 of 2006 (P=107.77). Note that all these bonds trade at a relatively modest premium, and therefore the systematic bond basis approximately coincides with the matching maturity curve basis.

Finally, we would like to note that comparing the BCDS and market CDS term structures may be a valuable tool not only for basis trading but for curve trading as well. For example, by looking at Figure 4, an investor may conclude that the CDS curve is too flat compared with the cash market-based BCDS curve. Depending on the investor's directional views regarding Altria, this opinion could be implemented in either bullish or bearish fashion.

A bearish curve steepener would be to sell 2-year protection and buy 7-year protection in equal notional amounts. Such a trade has an overall negative Credit01, and therefore will produce positive returns if spreads widen and/or steepen. A bullish curve steepener would be to sell 2-year protection and buy 7-year protection in amounts which make the trade Credit01-neutral. Then the overall spread moves will not affect the trade much, while the excess tightening in the front end as the spreads rally will produce positive returns.

**Figure 4. Curve and bond basis in Altria Group (as of 2/06/2004)**

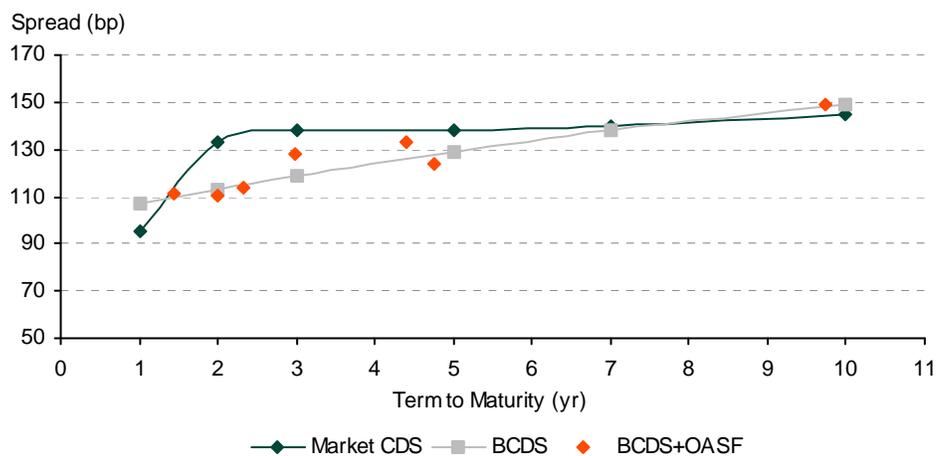

### THE COARSE-GRAINED HEDGING STRATEGY AND THE APPROXIMATE CDS-BOND BASIS MEASURES

The static hedging strategy using a sequence of forward CDS is difficult to implement in practice. Although less precise, the staggered strategy using spot CDS is generally easier to put to work. However, even the staggered strategy, if implemented in short term increments as presented in Figure 2b, would lead to odd-lot hedge notionals for intermediate terms, and likely result in an unacceptable loss of liquidity.

As a compromise between accuracy and liquidity, we suggest a coarse-grained staggered hedge which can be constructed using a maturity grid with longer intervals. The forward price changes in these intervals will yield lumpier intermediate hedge notionals, according to [6]. The optimal hedging grid will depend on the bond coupon level and the underlying interest rates. For bonds with modest premium or discount, just one or two additional hedges can result in sufficient accuracy.

Let us consider a single-CDS strategy first. Figure 5 shows an example of such strategy (the line with diamonds) contrasted with the theoretical precise strategy using forwards (the light solid line) in the case of a premium credit bond whose forward price (dashed line) gradually approaches par towards the 5-year maturity. Such a simple hedging strategy would typically be underhedged (compared to the theoretical requirement) during the early years and overhedged during the later years of its projected existence. The optimal hedge notional will the such that the net present value of the outstanding risk exposures (positive for short terms and negative for long terms) become precisely zero.

Such a requirement also helps in explaining why the "market hedge ratio" that is equal simply to the price premium of the underlying bond (which is quite popular among the practitioners, see McAdie and O'Kane [2001]) provides a good starting guess for the correct hedge amount. Indeed, had we approximated the theoretical hedge notional line by a straight line, and ignored the effect of interest rate discounting, the optimal notional would correspond to the mid-point between the final theoretical hedge ratio at maturity, i.e. 100%, and the initial theoretical hedge ratio at current time, i.e. $N = (P - R)/(1 - R)$ (see [4]). If one used a recovery rate of 50% which is quite close to the long-term average recovery estimates, one would obtain the hedge notional equal to the current price of the bond.

We can see therefore, that the market practice is not too different from the correct single-CDS optimal hedge. One must ask, however, whether the single-CDS hedge itself is the optimal solution, or are we missing something important by limiting ourselves to only one hedging instrument.

Consider now the two-CDS hedging strategy depicted in Figure 5 by stacked bars. The strategy consists of a face value hedge to final maturity, plus an additional (staggered) hedge to a shorter maturity. We can arrange this strategy to be, for example, overhedged during the earlier years and underhedged during the later years of its projected existence. The same requirement of zero net present value of the outstanding risk exposures will define the optimal notional of the smaller staggered hedge, given its chosen maturity.

Note also, that by allowing the maturity of the add-on staggered hedge to be equal to the final maturity of the bond we would recover the case of a single-CDS strategy. Therefore, the two-CDS strategy will always be at least as good as the single-CDS one. When will it be better? It would be a better choice when the cost of hedging using two CDS turns out to be less than the cost of hedging using a single CDS. In the case of a premium bond, this will be the case if the CDS spread term structure is upward sloping, making the use of shorter term CDS a preferable option.

**Figure 5. Coarse-grained hedging strategies with one and two CDS**

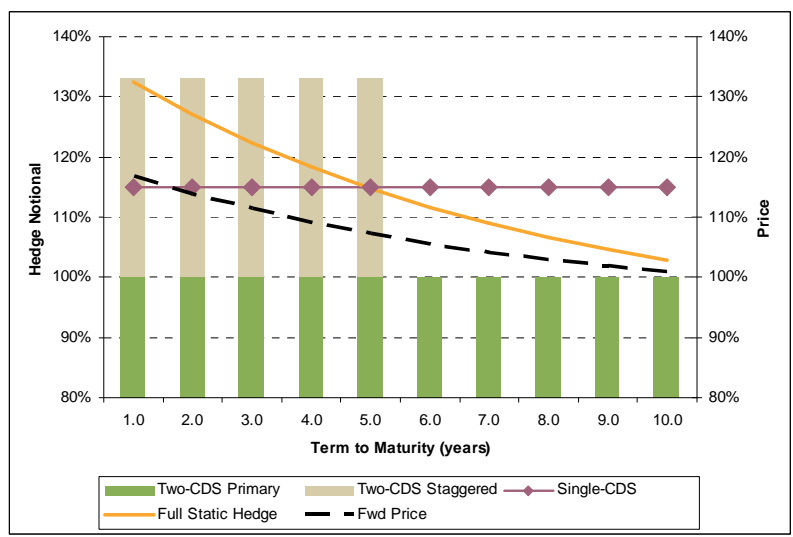

The considerations above lead us to the following simple recipe for a practical and accurate hedging strategy of credit bonds with CDS:

- Set the primary hedge to the final maturity of the bond with a notional equal to par face value of the bond.

- Consider various intermediate maturities (including the bond's final maturity) for which a sufficiently liquid market in CDS exists, and determine the optimal staggered hedge amount for the second CDS position for each maturity.

- Select the lowest cost hedging strategy among all considered two-CDS combinations.

The corresponding approximate CDS-Bond basis for the bond under consideration will be determined by comparison of its theoretical full BCDS spread (including the OAS-to-Fit correction) and the aggregate CDS spread of the coarse-grained staggered hedging strategy. This means that, even in a simplified framework, the correct relative value measure for basis trading strongly depends on the bond's coupon level and price premium, and the term structures of both interest rates and CDS spreads.

### OTHER RISKS AND DRIVERS OF THE BASIS

The consistent relative value measure for CDS-Bond basis considered in this article goes a good distance toward explaining the market-observed basis. Still, there are a number of additional both fundamental and technical reasons that affect the pricing of bonds and CDS and lead to the presence of the CDS-Bond basis even after correcting for the inherent biases associated with the commonly used Z-spreads or asset swap spreads. We list some of them below (see also McAdie and O'Kane [2001]).

Factors that drive CDS spreads wider than bonds:

- **Delivery option:** the standard CDS contract gives a protection buyer an option to choose a delivery instrument from a basket of deliverable securities in case of default.

- **Risk of technical default and restructuring:** the standard CDS contract may be triggered by events that do not constitute a full default or a bankruptcy of the obligor.

- **Delayed P&L realization:** in some cases, when the CDS position cannot be unwound with the same counterparty or assigned to another counterparty, the P&L realization will be delayed until maturity, and will also be subject to additional counterparty risk.

- **Demand for protection:** the difficulty of shorting credit risk in a bond market makes CDS a preferred alternative for hedgers and tends to push their spreads wider during the times of increasing credit risks.

- **LIBOR-spread vs. Treasury spread:** the CDS market implies trading relative to swaps curve, while most of the bond market trades relative to Treasury bond curve. Occasionally, the widening of the LIBOR spread that is driven by non-credit technical factors such as MBS hedging can make bonds appear "optically" tight to LIBOR.

Factors that drive CDS spreads tighter than bonds:

- **Implicit LIBOR-flat funding:** the CDS spreads imply a LIBOR-flat funding rate, which makes them cheap from the perspective of many protection sellers, such as hedge funds and lower credit quality counterparties, who normally fund at higher rates.

- **Counterparty credit risk:** the protection buyer is exposed to the counterparty risk of the protection seller and must be compensated by tighter CDS spreads.

- **Residual interest rate risk:** as we mentioned earlier in the paper, even after implementing the precise static hedging strategy which immunizes the expected price risk, there remains a residual interest rate risk contingent on the default of the issuer.

- **Differential accrued interest loss:** in the CDS market, the accrued interest is netted with the protection payment in case of default. In the bond market, the accrued coupon amount is often lost (or is considered at greater risk by the investors). We ignored the coupon-level effects in Appendix A when deriving the complementarity condition. However, the reader can see the origin of these effects from the more accurate approximation shown in the Appendix A of the Part 2 of this series.

- **Differential liquidity:** while the amount of the available liquidity in the top 50 or so bond issuers is greater in the cash market, the situation is often reverse for the rest of the credit market where writing protection can be easier than buying the bonds.

- **Synthetic CDO bid:** the large and growing market in synthetic CDOs is a source of demand for writing CDS protection which is driven at times by alternative selection criteria (ratings arbitrage, diversity rules, etc.) and may overwhelm the fundamentals encoded in the bond market (see Berd, Ganapati, Ha [2003]).

For all these reasons, the CDS-Bond basis can be and often is substantial. While the fundamental factors affect the proper value of the "fair" basis and are largely stable, the transient nature of the more powerful technical factors causes the basis to fluctuate around this fair value with a typical mean reversion times that ranges between a few weeks to months. This makes basis trading an attractive relative value investment strategy.

Many investors have been actively trading such basis convergence strategies by relying on the conventional measure, the difference between the CDS spread and the bond's LIBOR OAS (Z-spread). Figure 6 shows the historical estimates of the conventional CDS-Bond basis for 5-year term across a wide selection of investment grade issuers (see Berd [2004]). We can observe several significant regime changes during this period.

Some of these regimes reflect the prevalence of real driving forces, such as the loan hedging demand during 2002 and the synthetic CDO bid starting in late 2002 and continuing during the first half of 2003 (with a significant tempering of this demand in late spring after the spread compression made the ratings arbitrage much less attractive).

**Figure 6. Market-wide basis tone using conventional measures**

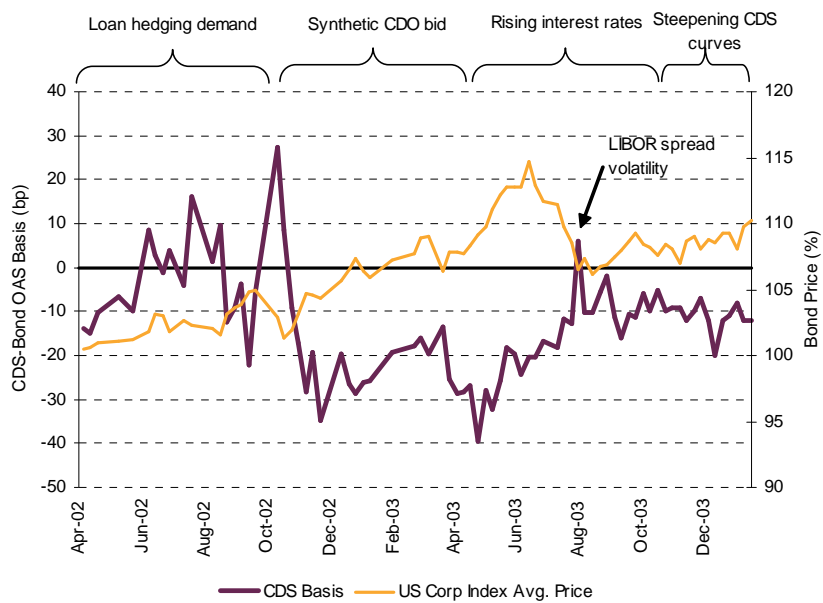

However, there have been other regimes that were driven by the inherent biases of the conventional CDS-Bond basis measures rather than real driving factors. Among these were the apparent convergence of the basis from negative to moderate levels during the summer 2003 that was driven by the rising interest rates and reduction in bond price premiums (illustrated in Figure 6 using the average price of Lehman Brothers U.S. Corporate Investment Grade Index). Latter months of 2003 were characterized by an uptick of price premiums and a significant steepening of the CDS curves, which was reflected in the apparent mild drift of the conventional basis measure towards negative values. Finally, in the beginning of August 2003 we witnessed wild short-term swings of LIBOR spreads driven by strong hedging pressures from the MBS market which resulted in an apparent blip on the CDS-Bond basis.

Understanding which of the regimes is prevalent, and whether the driving forces are real or "optical" would make a significant difference when designing and testing basis trading strategies. In particular, a strategy that bet on convergence of basis during the summer of 2003 was really an interest rate directional strategy and investors should evaluate it as such by asking a question on whether such an implementation of their views on rising interest rates using a CDS-Bond basis was more efficient compared to simply shorting the rates via swaps or Treasury futures.

The consistent estimates of the basis presented in this paper, and the understanding of its dependence on such driving factors as the interest rate and CDS spread curve shapes should help investors in both selecting the best relative value trades and in avoiding the "false positives", i.e. trades that seem to perform, but for reasons entirely different from the investor's original justification.

**CONCLUSIONS**

In the conclusion of our three-part series of papers we note that the survival-based pricing framework has important implications for all aspects of both the investment science and the practice of trading of credit-risky bonds. We have shown in this series that all commonly used measures, including the Z-spreads, the interest rate and spread durations, and the CDS-Bond basis, to name a few, are significantly affected and require a careful re-definition to maintain their consistency. The impact of the new methodology is both qualitative, such as the revealing explanations of the "optical" spread curve inversion and rates-spreads correlation, and quantitative, such as the estimates of the correct interest rate and credit sensitivities and CDS-Bond basis. We hope that the comprehensive set of methodologies presented in this series will assist the credit portfolio managers in the difficult task of understanding and harnessing the risks and rewards of this important asset class.

*Acknowledgements:* We would like to thank Marco Naldi as well as many other colleagues at Lehman Brothers Fixed Income Research department for numerous helpful discussions throughout the development and implementation of the survival-based methodology during the past several years.

## APPENDIX: THE COMPLEMENTARITY BETWEEN THE FORWARD CDS AND RISK-FREE-EQUIVALENT COUPON TERM STRUCTURES

Assume that the underlying risk-free discount curve $r(t)$ (usually Libor) and the issuer's hazard rate term structure $h(t)$ are given. The forward discount function for the riskless rate is given by a well known formula, where $r(s)$ is the instantaneous forward rate:

$$[11] \quad Z(t,T) = \exp\left(-\int_t^T r(s) \cdot ds\right)$$

The forward survival probability $Q(t,T)$ stands for the probability that the issuer will survive during the time period $(t,T)$ provided that it survived until the beginning of that period. It is related to the hazard rate by a similar relationship[1]:

$$[12] \quad Q(t,T) = \exp\left(-\int_t^T h(s) \cdot ds\right)$$

Consider a credit-risky bond with a given coupon $C$ and final maturity $T$. The projected forward price of a fixed coupon bond depends on both of these term structures as well as the level of the coupon in the following manner (for simplicity of exposition we use the continuous-time approximation and ignore the small corrections proportional to the coupon level – see Appendix A in Part 2 of this series for detailed derivation):

$$[13] \quad P(t,T) = C \cdot \int_t^T du \cdot e^{-\int_t^u (r(s)+h(s)) \cdot ds} + e^{-\int_t^T (r(s)+h(s)) \cdot ds} + R \cdot \int_t^T du \cdot h(u) \cdot e^{-\int_t^u (r(s)+h(s)) \cdot ds}$$

The first term reflects the present value of the coupon stream under the condition that the bond survived until some intermediate time $t$, the second term reflects the present value of the final principal payment under the condition that the bond survived until the final maturity $T$, the third term reflects the recovery of the fraction $R$ of the face value if the issuer defaults at any time between the valuation time and the final maturity.

Let us define the "risk-free-equivalent coupon" stream $RFC(t,T)$ which would reproduce the same forward price term structure but only when discounted with the risk-free discount function, without any default probability. Such a coupon stream will not be constant in general and will have a non-trivial term structure, depending on both the underlying risk-free rates and, through the price of the risky bond, on the issuer hazard rates as well. The defining condition is:

$$[14] \quad P(t,T) = \int_t^T du \cdot RFC(u,T) \cdot e^{-\int_t^u r(s) \cdot ds} + e^{-\int_t^T r(s) \cdot ds}$$

The concept of a risk-free equivalent coupon stream is necessary for consistent definition of the difference between the default-risky and risk-free bonds when the underlying interest and hazard rates have non-trivial term structures and the bonds are expected to deviate from par pricing either currently or at any time in the future.

---

[1] *The hazard rate and interest rate can, in general, be stochastic. In this discussion, we restrict our attention to deterministic default intensities and deterministic interest rates.*

To find the relationship between the risk-free-equivalent coupon stream $RFC(t,T)$ and the forward price $P(t,T)$ of the credit-risky bond, let us take a derivative with respect to the valuation time $t$ of both sides of equations [13] and [14]. On one hand we get:

[15] $\quad \dfrac{\partial P(t,T)}{\partial t} = (r(t) + h(t)) \cdot P(t,T) - C - R \cdot h(t)$

On the other hand, we get:

[16] $\quad \dfrac{\partial P(t,T)}{\partial t} = r(t) \cdot P(t,T) - RFC(t,T)$

Since the left-hand sides are equal by construction, we can equate the right-hand sides and obtain the relationship between the risk-free-equivalent coupon stream and the forward price:

[17] $\quad C - RFC(t,T) = h(t) \cdot (P(t,T) - R)$

Consider now a forward CDS contract for a short time period $(t, t + \Delta t)$. This is a contract which provides protection during the future time period in exchange for premiums which are paid during that period but whose level is preset today. In case of a credit event prior to starting date $t$, the forward CDS knocks out and provides no further protection during the future period. As explained in Berd (2003), the forward CDS spread is determined by:

[18] $\quad CDS_{fwd}(t,T) \cdot \int_t^T du \cdot e^{-\int_t^u (r(s) + h(s)) \cdot ds} = (1 - R) \cdot \int_t^T du \cdot h(u) \cdot e^{-\int_t^u (r(s) + h(s)) \cdot ds}$

It is easy to see that for a small future time interval, the forward spread is simply proportional to the hazard rate of the matching horizon:

[19] $\quad CDS_{fwd}(t, t + \Delta t) = (1 - R) \cdot h(t)$

Substituting this definition into equation [17], we get a complementarity condition between the risk-free-equivalent coupon streams and the forward CDS spreads:

[20] $\quad C - CDS_{fwd}(t) \cdot N(t,T) = RFC(t,T)$, where $N(t,T) = \dfrac{P(t,T) - R}{1 - R}$

This relationship confirms our intuition about the consistent hedging strategy for non-par credit-risky bonds which consists of a stream of forward CDS with notionals $N(t,T)$ depending on the forward price of the bond. The residual cashflows of the credit-risky bond after paying the required premiums coincide with the projected risk-free-equivalent coupon stream. Although there is still a timing risk associated with this hedging strategy, the notionals of the hedges are such that the recovered value will be equal to the correct forward price of the bond, and therefore the timing risk is unimportant when evaluating the present value of the hedged cashflows to the present time or to any future time before maturity. This is reflected in the fact that discounting these residual cashflows with riskless rate gives the correct forward prices of the bond (compare [14] and [20]).

Also note that the hedge notionals depend on the recovery rate both explicitly and implicitly, via the dependence of the implied hazard rates on the recovery rate. Therefore a credit bond hedged according to this strategy still contains recovery risk that may need to be hedged using either digital CDS or recovery swaps (see Berd and Kapoor [2002], and Berd [2005]).